\documentclass[twocolumn,showpacs,amsmath,amssymb,nofootinbib,amsfonts,showkeys]{revtex4}

\RequirePackage{graphicx}
\RequirePackage{amsmath}
\usepackage{multirow}
\usepackage{caption}

\begin{document}

\title{Constraints on the interaction of quintessence dark energy with dark matter and the evolution of its equation of state parameter}

\author{Roman Neomenko}
\email[]{roman.neomenko@lnu.edu.ua} 
\affiliation{Astronomical Observatory of Ivan Franko National University of Lviv, Kyryla i Mefodiya Street 8, Lviv, 79005, Ukraine}
\date{\today}

\begin{abstract}
The cosmological model with an interaction between dynamical quintessence dark energy and cold dark matter is considered. Evolution of a dark energy equation of state parameter is defined by a dark energy adiabatic sound speed and a dark sector interaction parameter, which must be more physically correct model then a previously used in which such evolution was given by some fixed dependence on scale factor. The constraints on interaction parameter and other parameters of the model was obtained using a cosmic microwave background, baryon acoustic oscillations and supernova SN Ia data.
\end{abstract}

\pacs{95.36.+x,95.35.+d,98.80.-k}

\keywords{interacting dark energy, dark matter, cosmological perturbations}

\maketitle

\section{Introduction}
\label{sec:intro}
The $\Lambda$CDM model is the most simple cosmological model which is in a very good agreement with astrophysical observational data. However a theoretical explanation of such model is very problematic from a quantum-field theoretical point of view, which is described in a reviews \cite{Carroll2001,Sahni2000}. The alternative to $\Lambda$ term in Einstein`s equations is the new component -- dark energy (DE) which in the most cosmological calculations is described as a perfect fluid with positive energy density and negative pressure which causes accelerated expansion of universe at present time. Such component can be easier explained theoretically: it can be for example some classical scalar field. The constraints on the equation of state parameter (EoS parameter) of DE, which in the most simple models is constant, point to EoS parameter value close to $-1$ (which is $\Lambda$CDM model) \cite{Planck2018}. And when considering the more general models of dynamical DE (in them the EoS parameter varies with the expansion of universe), their parameters constraints also prefer the $\Lambda$CDM model \cite{Planck2018}. So they don`t have advantage in the explanation of accelerated expansion of universe \cite{Riess1998,Perlmutter1999}. That's why there is a need to consider more complicated models of dynamical DE which could have a good theoretical explanation. One such are those, in which dynamical DE non-gravitationally interacts with cold dark matter (DM), which are called dynamical interacting dark energy models (dynamical IDE). This generalization is natural, because there are no known free fields in particle physics. These IDE models with constant or variable EoS parameter were studied in detail for various forms of interaction term in works, some of which are \cite{Amendola2000,Zimdahl2001,Amendola2007,Rui2018,Chimento2010,Rowland2008,Jesus2008,Gomez2020}. Anyway the previously studied dynamical IDE models \cite{Yang2018,Yang2023} have the problem of DE EoS parameter dependence on universe's expansion to be physically non-realistic. It is given by some fixed dependence on scale factor as for instance in \cite{Chevallier2001,Linder2003,Gong2005}. So we consider the model of dynamical IDE, in which evolution of the EoS parameter depends on internal properties of IDE. This evolution is parametrized by adiabatic sound speed of DE and a coupling parameter of DE-DM interaction \cite{Neomenko2016,Neomenko2020}. Also we consider the most widely used type of DE-DM interaction term, proportional to the energy density of DE \cite{Jackson2009,Gavela2009,DiValentino2017}. The problem with such types of interaction used in earlier works is that they are not general-covariant. Hence the general-covariant DE-DM interaction term \cite{Gavela2010} is used in our model of dynamical IDE. In this work for the first time the constraints on parameters of such model was obtained using Markov Chain Monte Carlo sampling method. It must be noted that here the parameters of the model is restricted to values where DE is quintessential. The model was compared with the cosmic microwave background (CMB), baryon acoustic oscillations (BAO) and supernova of type Ia (SN Ia) observational data.

This paper is organized as follows: in section II the general description of the model is given; in section III the observational data and analysis method, have been used in this work, are described; and in section IV the obtained results is discussed.

\section{Model of the dynamical IDE}
The description of dynamical IDE and all other components of the universe is made in perfect fluid approximation with stress-energy tensor:
\begin{equation}\label{tik}
T_{i}^{k}=(\rho+p)u_{i}u^{k}-p\delta_{i}^{k} \,.
\end{equation}
The universe is considered homogeneous and isotropic with zero spatial curvature, relative to which small perturbations of metrics and stress-energy tensor of each component occur (perturbations are given in synchronous gauge):
\begin{equation}\label{eq:ds2}
ds^{2}=a^{2}(\eta)[d\eta^{2}-(\delta_{\alpha\beta}+h_{\alpha\beta})dx^{\alpha}dx^{\beta}]\,,
\end{equation}
\begin{subequations}
\label{eq:est}
\begin{gather}
\bar{T}_{0}^{0}+\delta T_{0}^{0}=\bar{\rho}+\bar{\rho}\delta \,, \nonumber \\
\bar{T}_{0}^{\alpha}+\delta T_{0}^{\alpha}=0+(\bar{\rho}+\bar{p})v^{\alpha} \,, \nonumber \\
\bar{T}_{\alpha}^{0}+\delta T_{\alpha}^{0}=0-(\bar{\rho}+\bar{p})v^{\alpha} \,, \nonumber \\
\bar{T}_{\alpha}^{\beta}+\delta T_{\alpha}^{\beta}=-\bar{p}\delta_{\alpha}^{\beta}-\delta p\delta_{\alpha}^{\beta} \,. \nonumber
\end{gather}
\end{subequations}
Here $a$ denotes a scale factor, $\eta$ is conformal time, $h_{\alpha\beta}$ is perturbation of metric tensor, $\bar{\rho}$, $\bar{p}$ are background energy density and pressure, $v^{\alpha}\equiv\vec{v}\equiv dx^{\alpha}/d\eta$ is peculiar velocity, $\delta$ is perturbation of density relative to its background value, $\delta p$ is perturbation of pressure.

The sum of stress-energy tensors of dark components satisfy the general-covariant conservation equation, but separately due to DE-DM interaction they don't conserve:
\begin{subequations}\label{eq:tq}
\begin{gather}
\label{eq:tq:1}
T_{(de)i;k}^{k} = J_{(de)i} \,,
\\
\label{eq:tq:2}
T_{(c)i;k}^{k} = J_{(c)i} \,.
\end{gather}
\end{subequations}
Here $J_{i}$ is the DE-DM interaction term, $";"$ is a covariant derivative. From the conservation of total stress-energy tensor of dark components follows that $J_{(c)i}=-J_{(de)i}=J_{i}$. There are many forms of interaction term taken in works on IDE, but in this work one of the most often used is considered: its background zero component $\bar{J}_{0}$ is proportional to the energy density of DE \cite{Jackson2009,Gavela2009,DiValentino2017}:
\begin{equation}\label{j0de}
\bar{J}_{0}=3\beta aH\bar{\rho}_{de} \,.
\end{equation}
Here $\beta$ is the interaction parameter, $H\equiv(da/d\eta)/a^{2}$ is a Hubble parameter. This interaction form is popular because of the absence of DE perturbations' non-adiabatic instabilities at the radiation-dominated epoch of universe \cite{Jackson2009,Gavela2009}. The background component of equations \eqref{eq:tq} with this interaction take the following form:
\begin{subequations}\label{eq:rdedm}
\begin{gather}
\label{eq:rdedm:1}
\dot{\bar{\rho}}_{de}+3aH(1+w)\bar{\rho}_{de}=-3\beta aH\bar{\rho}_{de} \,,
\\
\label{eq:rdedm:2}
\dot{\bar{\rho}}_{c}+3aH\bar{\rho}_{c}=3\beta aH\bar{\rho}_{de} \,,
\end{gather}
\end{subequations}
where dot over $\bar{\rho}$ denotes its derivative on conformal time $\eta$ and $w$ is DE EoS parameter. In this work $w$ varies with the expansion of universe, hence the DE is dynamical. There are several parametrizations of $w$ evolution proposed in \cite{Chevallier2001,Linder2003,Gong2005}. In them EoS parameter is given as some function of scale factor $a$ which does not depend on the internal properties of DE. But it is obvious that the evolution of $w$ must depend on them, and in the case of our IDE model, on DE-DM interaction. To parametrize the evolution of $w$, which would satisfy these requirements, let's use adiabatic sound speed of DE. It's defined as $c_{a}^{2}=\dot{p}_{de}/\dot{\rho}_{de}$. Then from equation \eqref{eq:rdedm:1} one can be obtained the general equation for the evolution of $w$:
\begin{equation}
\label{eq:dw}
\frac{dw}{da}=\frac{3}{a}(1+w+\beta)(w-c_{a}^{2}) \,.
\end{equation}
In general case $c_{a}^{2}$ is dependent on time but in this work only the phenomenological models of IDE is considered for which $c_{a}^{2}=const$ \cite{Neomenko2016,Neomenko2020}. Such model of DE for non-interacting case was considered in \cite{Novosyadlyj2010,Novosyadlyj2012,Sergijenko2015}. The general solution of the system of equations \eqref{eq:rdedm}, \eqref{eq:dw} was obtained and analyzed in detail in work \cite{Neomenko2016}, and have the following form:
\begin{subequations}\label{eq:dedmw}
\begin{gather}
\label{eq:dedmw:1}
w=\frac{(1+c_{a}^{2}+\beta)(1+w_{0}+\beta)}{1+w_{0}+\beta-(w_{0}-c_{a}^{2})a^{3(1+c_{a}^{2}+\beta)}}-1-\beta \,,
\\
\label{eq:dedmw:2}
\bar{\rho}_{de}=\bar{\rho}_{de}^{(0)}\frac{(1+w_{0}+\beta)a^{-3(1+c_{a}^{2}+\beta)}-w_{0}+c_{a}^{2}}{1+c_{a}^{2}+\beta} \,,
\\
\label{eq:dedmw:3}
\bar{\rho}_{c}=\bar{\rho}_{c}^{(0)}a^{-3}+\beta\bar{\rho}_{de}^{(0)}\displaystyle\biggl[\biggl(\frac{A}{c_{a}^{2}+\beta}+B\biggr)a^{-3}- \nonumber \\
-\frac{A}{c_{a}^{2}+\beta}a^{-3(1+c_{a}^{2}+\beta)}-B\biggr] \,,
\\
A=\frac{1+w_{0}+\beta}{1+c_{a}^{2}+\beta}, \quad B=\frac{w_{0}-c_{a}^{2}}{1+c_{a}^{2}+\beta} \,, \nonumber
\end{gather}
\end{subequations}
where $w_{0}$, $\bar{\rho}_{de}^{(0)}$, $\bar{\rho}_{dm}^{(0)}$ are the values of EoS parameter, DE density and DM density at present time ($a=1$).

Also the extension of interaction term \eqref{j0de} to the background plus perturbation case is made as follows $J_{i}=3\beta H\rho_{de}u_{i}^{(c)}$ in works \cite{Jackson2009,Gavela2009}. This interaction form is not general-covariant. So in this work the interaction term is taken in such form:
\begin{equation}
J_{i}=3\beta\rho_{de}u^{(T)k}_{;k}u_{i}^{(c)} \,,
\end{equation}
where $u_{k}^{(T)}$ is the 4-velocity of the center of mass of all components in the universe. The presence of scalar quantity $u^{(T)k}_{;k}$ in this interaction term means that it takes into account the local deviations of the Hubble parameter from its background value $H$, and which was the first proposed in work \cite{Gavela2010}. Another general covariant form of $J_{i}$, which take into account the perturbations of $H$, but is not considered in this work, was proposed in \cite{Hoerning2023}.

The resulting equations for the cosmological perturbations of interacting dark components, following from \eqref{eq:tq}, in Fourier space, in synchronous gauge and in dark matter rest frame, take the such form \cite{Neomenko2020}:
\begin{subequations}\label{eq:prtb}
\begin{gather}
\dot{\delta}_{de}=-3aH(c_{s}^{2}-w)\delta_{de}-(1+w)\frac{\dot{h}}{2}-
\nonumber\\
-(1+w)[k^{2}+9a^{2}H^{2}(c_{s}^{2}-c_{a}^{2})]\frac{\theta_{de}}{k^{2}}-
\nonumber\\
-\beta\biggr[\frac{\dot{h}}{2}+\theta_{T}+9a^{2}H^{2}(c_{s}^{2}-c_{a}^{2})\frac{\theta_{de}}{k^{2}}\biggl]\,,\label{eq:prtb:1}
\\
\dot{\theta}_{de}=-aH(1-3c_{s}^{2})\theta_{de}+\frac{c_{s}^{2}k^{2}}{1+w}\delta_{de}+
\nonumber\\
+3aH\frac{\beta}{1+w}(1+c_{s}^{2})\theta_{de}\,,
\label{eq:prtb:2}
\\
\dot{\delta}_{c}=-\frac{\dot{h}}{2}-
\nonumber\\
-\beta\frac{\bar{\rho}_{de}}{\bar{\rho}_{c}}\biggr[3aH(\delta_{c}-\delta_{de})-\frac{\dot{h}}{2}-\theta_{T}\biggl]\,,
\label{eq:prtb:3}
\end{gather}
\end{subequations}
where $c_{s}^{2}$ is a comoving effective sound speed of DE, $\theta\equiv i(\overrightarrow{k},\overrightarrow{v})$ and
$$\theta_{T}=\frac{\sum_{N}(\bar{\rho}_{N}+\bar{p}_{N})\theta_{N}}{\sum_{N}(\bar{\rho}_{N}+\bar{p}_{N})}\,,$$
where $N$ denotes number of each component in the universe.

In this work the quintessence model of dynamical IDE is considered. So for the quintessence DE of our model, at the small values of scale factor $a\ll 1$, the EoS parameter is equal to the square of adiabatic sound speed $w_{0}=c_{a}^{2}$. Using this property of the EoS parameter evolution, the solutions of perturbation equations \eqref{eq:prtb} can be obtained at radiation-domination epoch and at supper-horizon scales ($k\eta<<1$):
\begin{subequations}
\label{eq:init}
\begin{gather}
\delta_{de}^{init}=\frac{3}{2}\frac{C}{E}\delta_{g}^{init}\,,\label{eq:init:1}
\\
\theta_{de}^{init}=18\frac{D}{E}\theta_{g}^{init}\,,\label{eq:init:2}
\\
\delta_{c}^{init}=\frac{3}{4}\delta_{g}^{init}\,,\label{eq:init:3}
\\
C=(1+c_{a}^{2}+\beta)[(4-3c_{s}^{2})(1+c_{a}^{2})- \nonumber\\
-3\beta(1+c_{s}^{2})]\,,\nonumber\\
D=c_{s}^{2}(1+c_{a}^{2}+\beta) \,, \nonumber\\
E=2(1+c_{a}^{2})(4+3c_{s}^{2}-6c_{a}^{2})- \nonumber\\
-3\beta(2+5c_{s}^{2}-3c_{a}^{2})\,, \nonumber
\end{gather}
\end{subequations}
where $\delta_{g}^{init}$, $\theta_{g}^{init}$ are initial perturbations of electro-magnetic radiation component. These solutions had been used as initial conditions when the integration of the system of perturbation equations was made.

\section{Observational data and method}

To obtain the observational constraints on the parameters of our model the following cosmological and astrophysical data were used:

{\bf 1. CMB temperature and polarization anisotropies.} The cosmological data on the anisotropies of the CMB, obtained by Planck Collaboration (Planck 2018 data release) \cite{CMB2018}. It contains the information on high-$l$ TT, TE, EE power spectra and low-$l$ TT and EE power spectra. Also in addition to this the data on the CMB weak gravitational lensing (Planck 2018 lensing) \cite{Lensing2018} were used.

{\bf 2. BAO data.} The following BAO observational data were used: SDSS-III Baryon Oscillation Spectroscopic Survey, DR12 \cite{DR12}; the 6dF Galaxy Survey \cite{6dF}; SDSS DR7 Main Galaxy Sample \cite{DR7}.

{\bf 3. SN Ia data.} Pantheon dataset \cite{Pantheon} which contains data on 1048 supernova of type Ia.

For constraining the dynamical IDE model parameters, a Markov Chain Monte-Carlo (MCMC) statistical method was used. For this the CosmoMC software package \cite{CosmoMC} was modified for our model. For the calculation of observable quantities of the model the CAMB code \cite{CAMB} was used, also modified for this purpose. The space of independent parameters has the three parameters $w_{0}$, $c_{a}^{2}$, $\beta$ in addition to the standard parameters of $\Lambda$CDM model. As the DE quintessence-phantom divide is shifted on $\beta$ \cite{Neomenko2016}, for the quintessence model the following conditions must be satisfied $w_{0},c_{a}^{2}>-1-\beta$. So it is convenient to introduce the renormalized quantities $\tilde{w}_{0}=w_{0}+\beta$ and $\tilde{c}_{a}^{2}=c_{a}^{2}+\beta$. Now for DE to be the quintessential the renormalized quantities must satisfy these conditions: $\tilde{w}_{0},\tilde{c}_{a}^{2}>-1$. For the interaction parameter $\beta$ the negative values of it are taken in the parameter priors, because at this values the DE cosmological perturbations are stable \cite{Neomenko2020}. Also the negative values of $\beta$ are preferred by MCMC constraints of quintessence IDE, made in the works \cite{DiValentino2017,DiValentino2020}. The resulting table of independent parameter priors taken in our MCMC model constraints are given in Table~\ref{tab:1}.

\begin{table}
\begin{center}
\caption{Priors for independent parameters.}  
\begin{tabular} {cc}
\hline
\hline
   \noalign{\smallskip}
Parameter&Prior  \\
 \noalign{\smallskip} 
\hline
   \noalign{\smallskip} 
$\Omega_{b}h^{2}$ & [0.005, 0.1] \\
$\Omega_{c}h^{2}$ & [0.001, 0.99] \\
$100\theta_{MC}$ & [0.5, 10] \\
$\tau$ & [0.01, 0.8] \\
$\log(10^{10}A_{s})$ & [1.61, 3.91] \\
$n_{s}$ & [0.8, 1.2] \\
$\tilde{w}_{0}$ & [-1, -0.333] \\
$\tilde{c}_{a}^{2}$ & [-1, 0] \\
$\beta$ & [-0.3, 0] \\ 
    \noalign{\smallskip}    
  \hline  
  \hline
\end{tabular}
\label{tab:1}
\end{center}
\end{table}

In the MCMC simulation the 8 chains were used. The Gelman-Rubin parameter, used as the measure of the chain convergence, is taken $R-1<0.01$ for the MCMC chains being converged.

\section{Results}

\begin{table}
\begin{center}
\caption{The parameter constraints of dynamical IDE model given at 68\% CL.}  
\begin{tabular}{cc}
\includegraphics[width=0.25\textwidth]{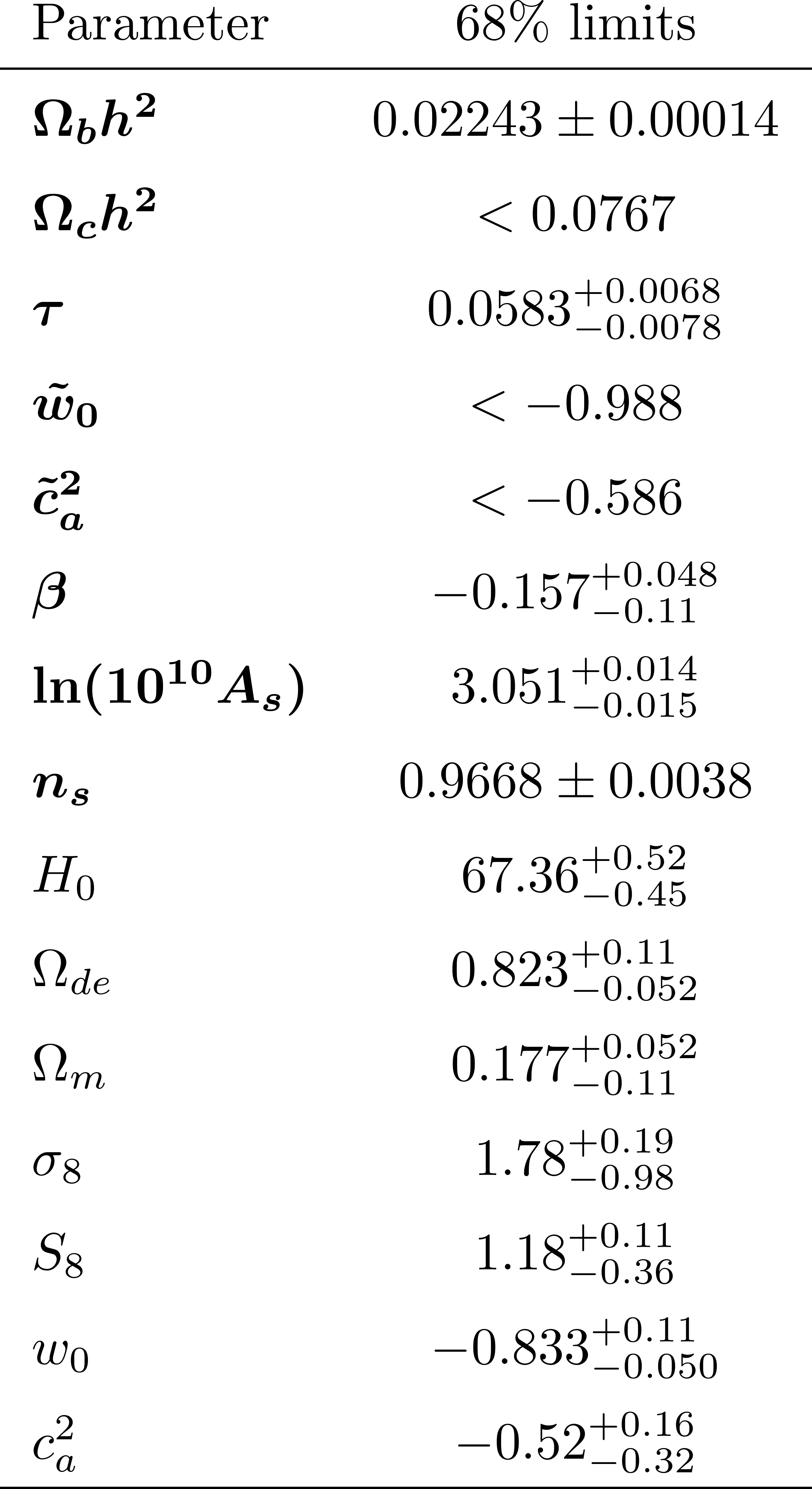}
\end{tabular}
\label{tab:2}
\end{center}
\end{table}

\begin{figure}
\centering
\includegraphics[width=0.3\textwidth]{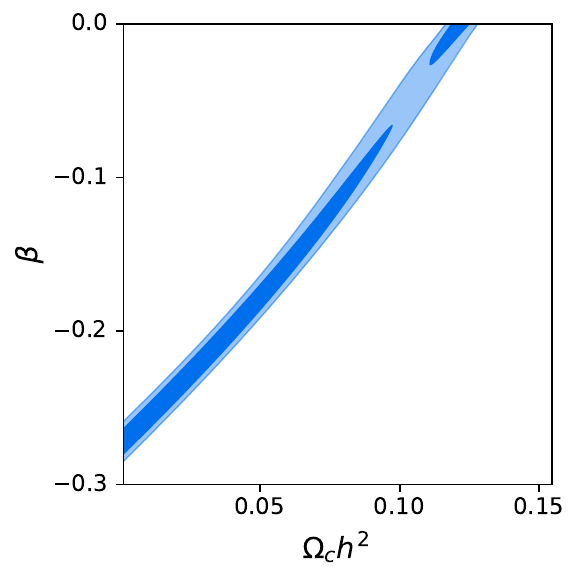}
\caption{The 2D marginalized distribution in $\Omega_{c}h^{2}$--$\beta$ plane of the dynamical IDE model.}
\label{fig:1}
\end{figure}

The results on the MCMC constraining of dynamical IDE model parameters are given in Table~\ref{tab:2} for 68\% CL limit. The comparison of the model with observational data prefer the non-zero value of interaction parameter $\beta$ at $2.05\sigma$ significance level. Also there is degeneracy between $\Omega_{c}h^{2}$ and $\beta$ parameters as it is seen from  the 2D-marginalized distribution of $\Omega_{c}h^{2}$--$\beta$ parameters shown in Fig.~\ref{fig:1}. This occurs because the amount of DM in the universe is directly dependent on the energy transfer rate from DM to DE. This fact doesn't alow to determine the $\Omega_{c}h^{2}$ lower bound in the considering model, using Planck 2018+lensing, BAO and Pantheon datasets only. In work \cite{DiValentino2020} for a $\xi q$CDM model (with the interaction of form $J_{i}=\xi H\rho_{de}u_{i}^{(c)}$) such degeneracy also occurred, with the absence of lower bound constraint on $\Omega_{c}h^{2}$. For the same interaction in the quintessence IDE model (but with the priors of $\xi$ bounded to the positive values), considered in work \cite{Yang2020} and the vacuum IDE model, considered in work \cite{Hoerning2023} (Model IV) and in work \cite{DiValentino2020} (model $\xi\Lambda$CDM) the degeneracy between these parameters is also present, when constraining model parameters using the Planck data only, and breaks down when adding additional datasets. So it is expected that adding other observational data which wasn't used in the MCMC parameter constraints in our work would break this $\Omega_{c}h^{2}$--$\beta$ degeneracy and would give the tighter bounds on the dynamical IDE model parameters. This behaviour of IDE model can also be different for the other forms of interaction $J_{i}$, such as for Model II and Model III in \cite{Hoerning2023}, where correlation between $\Omega_{c}h^{2}$ and $\beta$ is much smaller. The negative non-zero value of $\beta$ means that in the future epochs of universe the energy transfer from DM to DE will lead to negative values of the DM energy density $\rho_{c}$.

Also, the upper bound of DE EoS parameter at present time $w_{0}$ overlaps with the lower bound of the square of DE adiabatic sound speed $c_{a}^{2}$ as it is shown in Table~II. Hence we can`t determine with high significance level whether $w$ in our model varies with the expansion of universe.

\section{Conclusions}

In this work a constraints on a model parameters of dynamical interacting quintessence dark energy was made. Unlike in the previous works on this kind of cosmological models, constraints were made for the first time for the model, in which the coupling in dark sector has the general-covariant form and the evolution of dark energy equation of state parameter is dependent on the internal properties of dark energy including its coupling with dark matter. From the results of parameter constraining using CMB, BAO and SN Ia observational data follows the non-zero value of coupling parameter at $2.05\sigma$ significance level. However the constraints on the evolution of the dark energy equation of state parameter are not very tight, so it is uncertain whether the dark energy is dynamical at all. Also there is degeneracy between the amount of the dark matter in the universe and interaction strength in dark sector, which doesn't alow to obtain the lower bound on $\Omega_{c}h^{2}$ parameter. It is expected that using additional observational data in the future statistical analysis of this model will give more precise constraints on its parameters.

\section{Acknowledgements}

This work was supported by the project of the Ministry of Education and Science of Ukraine "Modeling the luminosity of elements of the large-scale structure of the early universe and the remnants of galactic supernovae, the observation of variable stars" (state registration number 0122U001834) and by the National Research Foundation of Ukraine under Project No. 2020.02/0073.

\end{document}